\documentclass[sigconf]{acmart}

\usepackage{enumitem}
\usepackage{listings}
\usepackage{parcolumns}
\usepackage{xcolor}

\definecolor{codegreen}{rgb}{0,0.6,0}
\definecolor{codegray}{rgb}{0.5,0.5,0.5}
\definecolor{codepurple}{rgb}{0.58,0,0.82}
\definecolor{backcolour}{rgb}{0.95,0.95,0.92}

\lstdefinestyle{mystyle}{
    commentstyle=\color{codegreen},
    keywordstyle=\color{magenta},
    numberstyle=\tiny\color{codegray},
    stringstyle=\color{codepurple},
    basicstyle=\ttfamily\footnotesize,
    breakatwhitespace=false,         
    breaklines=true,                 
    captionpos=b,                    
    keepspaces=true,                 
    numbers=left,                    
    numbersep=5pt,                  
    showspaces=false,                
    showstringspaces=false,
    showtabs=false,                  
    tabsize=2
}

\AtBeginDocument{%
  \providecommand\BibTeX{{%
    \normalfont B\kern-0.5em{\scshape i\kern-0.25em b}\kern-0.8em\TeX}}}

\setcopyright{acmcopyright}
\copyrightyear{2022}
\acmYear{2022}
\acmDOI{10.1145/1122445.1122456}

\acmConference[MSR '22]{MSR '22: The 2022 Mining Software Repositories Conference}{May 23--24, 2022}{Pittsburgh, PA}
\acmBooktitle{MSR '22: The 2022 Mining Software Repositories Conference,
  May 23--24, 2022, Pittsburgh, USA}
\acmPrice{15.00}
\acmISBN{978-1-4503-XXXX-X/18/06}

\begin{document}

\title{Recommending Code Improvements Based on Stack Overflow Answer Edits}
\titlenote{The study was accepted at the MSR 2022 Registered Reports Track.}

\author{Chaiyong Ragkhitwetsagul}
\authornote{Both authors contributed equally to this research.}
\email{chaiyong.rag@mahidol.edu}
\affiliation{%
  \institution{Facutly of ICT, Mahidol University}
  \streetaddress{street}
  \city{Salaya}
  \state{Nakhon Pathom}
  \country{Thailand}
  \postcode{postcode}
}

\author{Matheus Paixao}
\email{matheus.paixao@uece.br}
\affiliation{%
  \institution{State University of Ceara (UECE)}
  \streetaddress{street}
  \city{Fortaleza}
  \state{Ceará}
  \country{Brazil}
}

\renewcommand{\shortauthors}{Ragkhitwetsagul and Paixao}

\begin{abstract}
\textbf{Background:} Sub-optimal code is prevalent in software systems.
Developers may write low-quality code due to many reasons, such as lack of technical knowledge, lack of experience, time pressure, management decisions, and even unhappiness.
Once sub-optimal code is unknowingly (or knowingly) integrated into the codebase of software systems, its accumulation may lead to large maintenance costs and technical debt.
Stack Overflow is a popular website for programmers to ask questions and share their code snippets.
The crowdsourced and collaborative nature of Stack Overflow has created a large source of programming knowledge that can be leveraged to assist developers in their day-to-day activities.

\noindent\textbf{Objective:} 
In this paper, we present an exploratory study to evaluate the usefulness of recommending code improvements based on Stack Overflow answers' edits.

\noindent\textbf{Method:}
We propose \textsc{Matcha}, a code recommendation tool that leverages Stack Overflow code snippets with version history and code clone search techniques to identify sub-optimal code in software projects and suggest their optimised version.
By using SOTorrent and GitHub datasets, we will quali-quantitatively investigate the usefulness of recommendations given by \textsc{Matcha} to developers using manual categorisation of the recommendations and acceptance of pull-requests to open-source projects.
\end{abstract}



\maketitle

\section{Introduction}

Sub-optimal code is prevalent in software systems, where developers may write low-quality code due to many reasons, such as lack of technical knowledge~\cite{Shrestha2020}, lack of experience~\cite{Eyolfson2011}, time pressure~\cite{Kuutila2020}, management decisions~\cite{Lavallee2015,Freire2020}, and even unhappiness~\cite{Graziotin2017}.
Regardless of the reason behind sub-optimal code, it can unknowingly (or knowingly~\cite{Bavota2016}) be integrated into the codebase of software systems.
The accumulation of sub-optimal code without proper remediation or prevention may lead to large maintenance costs~\cite{Sjoberg2013} and technical debt~\cite{Tom2013}.

Stack Overflow\footnote{\url{https://stackoverflow.com/}} is a popular website for programmers to ask questions and share their code snippets. 
The collaborative nature of Stack Overflow has created a large source of programming knowledge.
As a result, programmers commonly reuse code from answers in Stack Overflow in their software projects~\cite{Yang2017,Ragkhitwetsagul2019,Zhang2019}. 
Although code snippets from Stack Overflow may contain issues such as security vulnerabilities~\cite{Acar2016}, API misuses~\cite{Zhang2018}, and license-violating code~\cite{Ragkhitwetsagul2019,Zhang2019}, the crowdsourcing nature of the website allows for constant community updates that not only optimise the answers' code but also fix potential issues~\cite{Tang2021,Diamantopoulos2019}.

Several tools have been created that use the knowledge on Stack Overflow to assist developers. 
The tasks for which the tools were created are varied, such as providing working code examples~\cite{Keivanloo2014}, showing relevant Stack Overflow posts according to the code context in the IDE~\cite{Ponzanelli2013,Ponzanelli2014}, or improving API documentation~\cite{Treude2016}. 
Following this track, in this paper, we present an exploratory study to evaluate the usefulness of recommending code improvements based on Stack Overflow answers’ edits.
However, before we detail our proposal, we present an example from a Stack Overflow post.

\subsection{Motivating Example}
\label{subsec:motivating_example}

As a motivating example for this exploratory study, we take a close look at the Stack Overflow question (and its answers) number 40665315\footnote{\url{https://stackoverflow.com/questions/40665315}}.
In the question, the developer is using the Rest Assured\footnote{\url{https://rest-assured.io/}} library to assist in the writing of unit tests for a REST application developed with the Spring Boot\footnote{\url{https://spring.io/projects/spring-boot}} framework. 
For this particular unit test, the developer needs to set the port the service will be running from, which is the main topic of the question.

For this question, the same developer who asked the question figured out the solution and posted an accepted answer.
The code snippet for the answer is displayed in Listing~\ref{list:original_code}.
The solution found by the developer can be seen in Line 10, where a method \texttt{.port()} is used to set the service's running port and test its status code.
This solution was posted in November 2016, and this code was displayed on Stack Overflow as the accepted answer for question 40665315.
After some time, another answer was posted for this question by a different user.
In the alternative answer, the answerer argues that the unit test in the accepted answer was setting up the port every time the test was executed, which was a waste of resources because, if the class had multiple tests, each test would need to set the port again.
To solve this, the answerer proposed the creation of a \texttt{setUp()} method that would set the port only once for all the tests in the class.
Hence, in December 2018, the accepted answer was edited according to the suggestion, and its current code is displayed in Listing~\ref{list:latest_code}.
As one can see from Lines 8 to 11, the new \texttt{setUp()} method was added, and the call for method \texttt{.port()} (Line 15) in the test was removed.

\begin{table*}
  \noindent\begin{minipage}{1\columnwidth}
    \begin{lstlisting}[language=Java, caption=Code from the original accepted answer to Stack Overflow question number 40665315, label={list:original_code}]
    @RunWith(SpringRunner.class)
    @SpringBootTest(webEnvironment = SpringBootTest.WebEnvironment.RANDOM_PORT)
    
    public class SizesRestControllerIT {
        @LocalServerPort
        int port;
    
        @Test
        public void test2() throws InterruptedException {
            given().port(port).basePath("/clothes").get("").then().statusCode(200);
        }
    }
    
    
    
    
    \end{lstlisting}
  \end{minipage}\hfill
  \begin{minipage}{1\columnwidth}
    \begin{lstlisting}[language=Java, caption=Code from the updated accepted answer to Stack Overflow question number 40665315, label={list:latest_code}]
    @RunWith(SpringRunner.class)
    @SpringBootTest(webEnvironment = SpringBootTest.WebEnvironment.RANDOM_PORT)
    
    public class SizesRestControllerIT {
        @LocalServerPort
        int port;

        @Before
        public void setUp() {
            RestAssured.port = port;
        }

        @Test
        public void test2() throws InterruptedException {
            given().basePath("/clothes").get("").then().statusCode(200);
        }
    }\end{lstlisting}
  \end{minipage}
\end{table*}

From this example, one can see that the initial solution provided as the accepted answer, albeit fully functioning, was sub-optimal.
Despite the snippet in today's answer being an optimised version of the solution, the sub-optimal solution was displayed as the accepted answer for two years.
During this time, all Stack Overflow users who viewed this particular question would use the sub-optimal solution as the inspiration for their solutions.
Moreover, the optimisation needed to prevent the resource waste is so subtle that it would not be surprising that other developers would achieve a similar sub-optimal solution and never realise its potential problems.

For a quick assessment of this assumption, we took the statement causing the sub-optimal behaviour in the first version of the answer (\texttt{given().port(port)}), and searched GitHub's search service\footnote{\url{https://github.com/search}} with the statement as a string and filtering for Java projects.
At the time of writing this paper, this query returned 435 results in GitHub's search.
By looking at the 10 first results, we observed that 5 of them had no set up method and the port had to be set for each and every test being executed.
The average number of tests in these classes was 16.
In 2 out of the 10 results, there was a set up method but the port set up was not included as one of the set up steps.
Finally, in 3 out of the 10 results, there was a set up method with a proper port set up, as suggested in the optimised solution.

As one can see from this example, sub-optimal solutions are posted (and accepted) on Stack Overflow, and it may take a while until the solution is optimised by the community. 
In the meantime, developers may use the sub-optimal snippets in their own solutions.
In some cases, such sub-optimal solutions may be reached by developers on their own and not necessarily be copied from Stack Overflow.
Regardless of the reason for the sub-optimal code snippets, the large crowdsourced programming knowledge on Stack Overflow can be leveraged to benefit developers.

\subsection{Study's Proposal and Research Questions}

In this paper, we propose an exploratory study of our approach, called \textsc{Matcha}, to recommend improvements to sub-optimal code based on answer edits on Stack Overflow. 
According to the motivating example, we can see that code snippets in a Stack Overflow accepted answer evolve over time and some of the accepted answers are later updated to include code improvements. 
\color{black}
\textsc{Matcha} operates in two main steps: 1) \textsc{Matcha} searches for similar code between a software project's code and snippets in Stack Overflow answers, 2) \textsc{Matcha} recommends the latest version of the answer to developers.
\color{black}
The approach leverages the scalable code clone search technique of Siamese~\cite{Ragkhitwetsagul2019} to retrieve similar code snippets from a large corpus of Stack Overflow accepted answers. 
We aim to augment Siamese to handle multiple revisions of the same code snippet, which is the case in Stack Overflow answers. 
Importantly, we will include the recommendation module in Siamese to be able to return the latest revision of the code snippet as the search result. 
We foresee that this approach will be useful for developers who are unknowingly using sub-optimal snippets in their software. 
They can adopt the recommendation to improve the quality of their implementation. 
Lastly, we designed this exploratory study to evaluate the usefulness of the recommendations given by our approach.

In the study, we intend to answer the following research questions:
\begin{itemize}[leftmargin=*]
\color{black}
\item \textbf{RQ1:} \textit{How accurate is \textsc{Matcha}'s detection of similar code snippets between software projects and Stack Overflow answers?} \textbf{(Sanity Check)}

\textbf{Objective:}~To check \textsc{Matcha}'s accuracy regarding the detection of similar code snippets between software projects and Stack Overflow answers. 
The combination of Siamese and the additional modules (see Section~\ref{sec:design}) compose the code clone search engine employed by \textsc{Matcha}. 
Hence, this serves as a sanity check for our exploratory study because it evaluates the core underlying technology behind \textsc{Matcha}.

\textbf{Evaluation:}~We will employ the established code clone ground truth between GitHub projects and Stack Overflow posts provided by the SOTorrent dataset~\cite{Baltes2018a}. By leveraging this dataset, we will extract the 69,885 Java method clone pairs between GitHub and Stack Overflow to evaluate the accuracy of \textsc{Matcha}. 
We will search for the optimised configurations of Siamese and also compare the accuracy before and after integrating the \textit{boiler-plate code filter} module required by \textsc{Matcha} (explained in detail in Section~\ref{sec:design}).
\color{black}

\item \textbf{RQ2:} \textit{To what extent are the recommendations given by \textsc{Matcha} useful for developers?}

\textbf{Objective:}~To assess how often \textsc{Matcha} provides recommendations with relevant code updates that are useful for developers.

\textbf{Evaluation:}~We will consider our own dataset of GitHub projects selected for this exploratory study (see Section~\ref{subsec:phase2}). 
For each project in the dataset, we will run \textsc{Matcha} and store its recommendations.
Next, the recommendations will be manually classified according to Baltes et. al's~\cite{Baltes2020} categorisation of Stack Overflow post edits. 
Finally, the recommendations that are classified as relevant code updates (e.g., \texttt{Optimising} and \texttt{Refactoring}) will be considered useful for developers.

\item \textbf{RQ3:} \textit{To what extent are \textsc{Matcha}'s recommendations accepted in practice?}

\textbf{Objective:}~To assess the potential of \textsc{Matcha} being adopted by real-world developers and integrated into their development practices.

\textbf{Evaluation:}~We will select a subset of the recommendations considered useful in RQ2.
For each of the selected recommendations, we will open a pull-request to the GitHub project containing the code change and a description of the change according to the original Stack Overflow post.
The outcome of the pull-requests alongside the discussions between developers will be used to quali-quantitatively answer this research question.
\end{itemize}

\section{Background and Related Work} 

\subsection{Recommending Code Snippets}

In a recent study, Tang et al.~\cite{Tang2021} proposed a method to automatically identify comment-edit pairs on Stack Overflow, i.e., comments in an answer that trigger an edit in the answer.
One of the usage scenarios for the comments-edit pairs envisioned by the authors is the recommendation of the edits to projects with the same outdated snippet, where the comment would be used as the natural language description of the recommendation.
The approach proposed in this related work presents a few shortcomings. 
First, the answers' updates considered in the related work are limited to the ones triggered by comments, which excludes all other answers' edits due to other reasons, such as the one presented in our motivating example (see Section~\ref{subsec:motivating_example}).
Differently, \textsc{Matcha} considers all answers edits on Stack Overflow as candidates for recommendation, enlarging its pool of optimised snippets.
Second, in the related work, only exact matches (Type-1 clones~\cite{Roy2009}) between snippets were considered for recommendation.
By leveraging Siamese's ability to search for Type-1, Type-2 and Type-3 clones~\cite{Roy2009,Ragkhitwetsagul2019}, \textsc{Matcha} expands its ability to find sub-optimal snippets in software projects.


\color{black}

The papers by Ponzanelli et al.~\cite{Ponzanelli2013,Ponzanelli2014} present \textsc{Prompter}, an IDE plugin to recommend Stack Overflow discussions based on the developer's current coding context.
By leveraging existing search engines, \textsc{Prompter} identifies relevant discussions based on a predefined threshold.
The developer working in the IDE is given the possibility to open and visualise the Stack Overflow discussion directly in the IDE.
Despite not directly recommending code snippets to developers, \textsc{Prompter} leverages Stack Overflow's knowledge to enhance the developer experience in the IDE by providing more context to its coding decisions.
On a similar note, the work by Rubei et al.~\cite{Rubei2020} sets out to recommend relevant posts based on the developer's coding context.
The proposed tool, called \textsc{PostFinder} focuses on enhancing both Stack Overflow posts and the developer's code with additional metadata to boost the matching.
The results indicate how the new features enable the matching of code context to highly relevant posts.
The work by Rahman et al.~\cite{Rahman2016} proposes \textsc{RACK}, an automated tool focused on API recommendation from Stack Overflow knowledge.
Different from the previously mentioned work, the \textsc{RACK} tool leverages natural language queries to search for relevant Stack Overflow posts.

\color{black}

The community working on code recommendations has been rapidly growing, with papers proposing creative ways to recommend code from sources other than Stack Overflow.
Aroma~\cite{Luan2019} is a tool that offers code recommendations based on structural code search. 
It compares the code query's parse tree to code snippets in the index, prune the results, and cluster the remaining results to give high-quality recommendations. 
Keivanloo et al.~\cite{Keivanloo2014} proposes an approach to find working code examples from a large code corpus on the Internet by using maximal frequent itemset mining and custom search ranking function. 
Nyamawe et al.~\cite{Nyamawe2018} recommend refactoring solutions by using traceability and code metrics. 


\subsection{SOTorrent} 
\label{subsec:sotorrent}
SOTorrent~\cite{Baltes2018a} is the largest dataset of Stack Overflow data to date. The dataset is created from the Stack Overflow data dump augmented with the version history of Stack Overflow content. The version history can be retrieved at the level of whole post or individual post block. It also contains references of GitHub files to Stack Overflow posts. The dataset is periodically updated and the latest version is from December 2020 (version 2020-12-31) containing the content of 51,296,931 Stack Overflow posts with 81,536,422 post versions. From our initial analysis, there are 31,659 Java accepted code answers with revisions. The dataset can be accessed via Google BigQuery\footnote{\url{https://console.cloud.google.com/bigquery?project=sotorrent-org}} or Zenodo\footnote{\url{https://zenodo.org/record/3746061}}. 
\textsc{Matcha} uses SOTorrent as a source of code snippets to recommend code changes.

\subsection{Siamese}
\label{subsec:siamese}

Siamese (Scalable and Incremental Code Clone Search via Multiple Code Representations)~\cite{Ragkhitwetsagul2019} is a novel code clone search technique that is accurate and scalable to hundreds of million lines of code. The technique includes multiple code representations by transforming code into various representations to capture different types of code clones, query reduction that keeps only highly relevant terms in the code query, and a customised ranking function that allows selection of a preferred clone type to be ranked first. 
It offers accurate clone search with high precision and recall for Type-1 to Type-3 clones compared to other state-of-the-art code search and code clone detection tools~\cite{Ragkhitwetsagul2019}. 
Siamese's clone search architecture leverages the inverted index to efficiently search for code clones. It scales to a large code corpus by indexing 365M lines of code in less than a day. Each query response time is within 8 seconds. The technique is general and can be applied to several software engineering problems such as retrieving similar code snippets from a large-scale codebase, finding code clones between Stack Overflow and GitHub projects, and analysing software license violations. Currently, Siamese supports code clone search in Java language only. However, the technique can be extended to other languages by adding additional language parsers.

\color{black}
\section{\textsc{Matcha}'s Design}
\label{sec:design} 
\textsc{Matcha} is a code recommendation system that accepts a Java project as input and returns a list of code recommendations, i.e., snippets from Stack Overflow answers that are similar to the input snippet with the latest updates. 
\textsc{Matcha}'s main component is the Siamese code clone search engine~\cite{Ragkhitwetsagul2019}. 
For \textsc{Matcha}, we will augment Siamese to include three additional modules: \textit{boiler-plate code filter}, \textit{multiple-code revision search}, and \textit{latest code revision retrieval} (see Figure~\ref{fig:overview}). 
The three modules are crucial for giving high-quality code recommendations by \textsc{Matcha}. 
Although the input is given to \textsc{Matcha} as a project, \textsc{Matcha} will read each Java file in the project, parse them, and perform the code clone search at method-level one at a time. 
During the clone search process of each given code query, \textsc{Matcha} will perform the search using Siamese by applying multiple code representation, query reduction, and clone ranking. Within the search process, we will incorporate the three additional modules to enable \textsc{Matcha} to return code recommendations as follows.
Siamese is currently written in the Java language.
Similarly, \textsc{Matcha} and its new modules will also be implemented in Java.

\textbf{Boiler-plate Code Filter:}~
Clones between Stack Overflow and software projects can contain a large number of boiler-plate code~\cite{Ragkhitwetsagul2019a} (e.g., getters, setters, or \texttt{equals} methods) that are not useful for developers. 
Thus, the \textit{boiler-plate code filter} removes boiler-plate code from the search. 
The filter will be created as follows. 
First, we will compile a list of code patterns that are considered as boiler-plate code according to the classification in the previous study~\cite{Ragkhitwetsagul2019a}. 
For example, the \texttt{getter}, \texttt{setter}, \texttt{equals()}, \texttt{compareTo()}, \texttt{toString()} methods will be considered as boiler-plate code. 
Second, we will create a list of regular expressions that match with such boiler-plate code patterns. These regular expressions will be integrated into the search component of Siamese as a query filter. We will check the given code query against the list of regular expressions. If there is a match, we will skip that code snippet from performing the search, thus removing the boiler-plate code from the search results.

\textbf{Multiple Code Revision Search:}~
This module allows \textsc{Matcha} to search for multiple revisions of the same code snippet in a Stack Overflow answer. The \textit{multiple code revision search} module will be created as follows. 
First, we will create the clone search index of Siamese by inputting all the revisions of each Stack Overflow accepted answer in Java. We will name the code snippets in each revision by concatenating the \texttt{PostID} (the unique ID of each answer), the \texttt{LocalID} (the unique ID of each code block within the code) and the \texttt{HistoryID} (the unique ID of each revision of the code block) as defined by the SOTorrent dataset. For example, the Stack Overflow answer ID 8394534 has 3 revisions\footnote{\url{https://stackoverflow.com/posts/8394534/revisions}}, they will be indexed as follows. The code snippet in the original version will be saved into a file and indexed using the file name \texttt{8394534\_0\_0.java}. The first revision version will be created and indexed as \texttt{8394534\_0\_1.java}. The second revision, the latest one, will be created and indexed as \texttt{8394534\_0\_latest.java}. Using this naming technique, we can automatically identify from the Siamese clone search results if the matched code snippet is the latest version or not by checking from the name of the file name of the first-ranked result.

\textbf{Latest Code Revision Retrieval:}~
The \textit{latest code revision retrieval} module allows \textsc{Matcha} to return the latest revision of the matched code snippet. Based on the result from the \textit{multiple code revision search} module, \textsc{Matcha} will check, for each code snippet query, whether the latest version of Stack Overflow answer is returned. If not, \textsc{Matcha} will include the latest version of the answer into the list of code recommendations. 

After finishing searching using all the methods in the given Java project, \textsc{Matcha} will return the list of code recommendations in a CSV format. Each record contains the file name, method name, start line, and end line of the code in the project, and the \texttt{PostID} of the Stack Overflow post that contains the latest code revision.
\color{black}

\section{Execution Plan}

\begin{figure*}
  \includegraphics[width=0.9\textwidth]{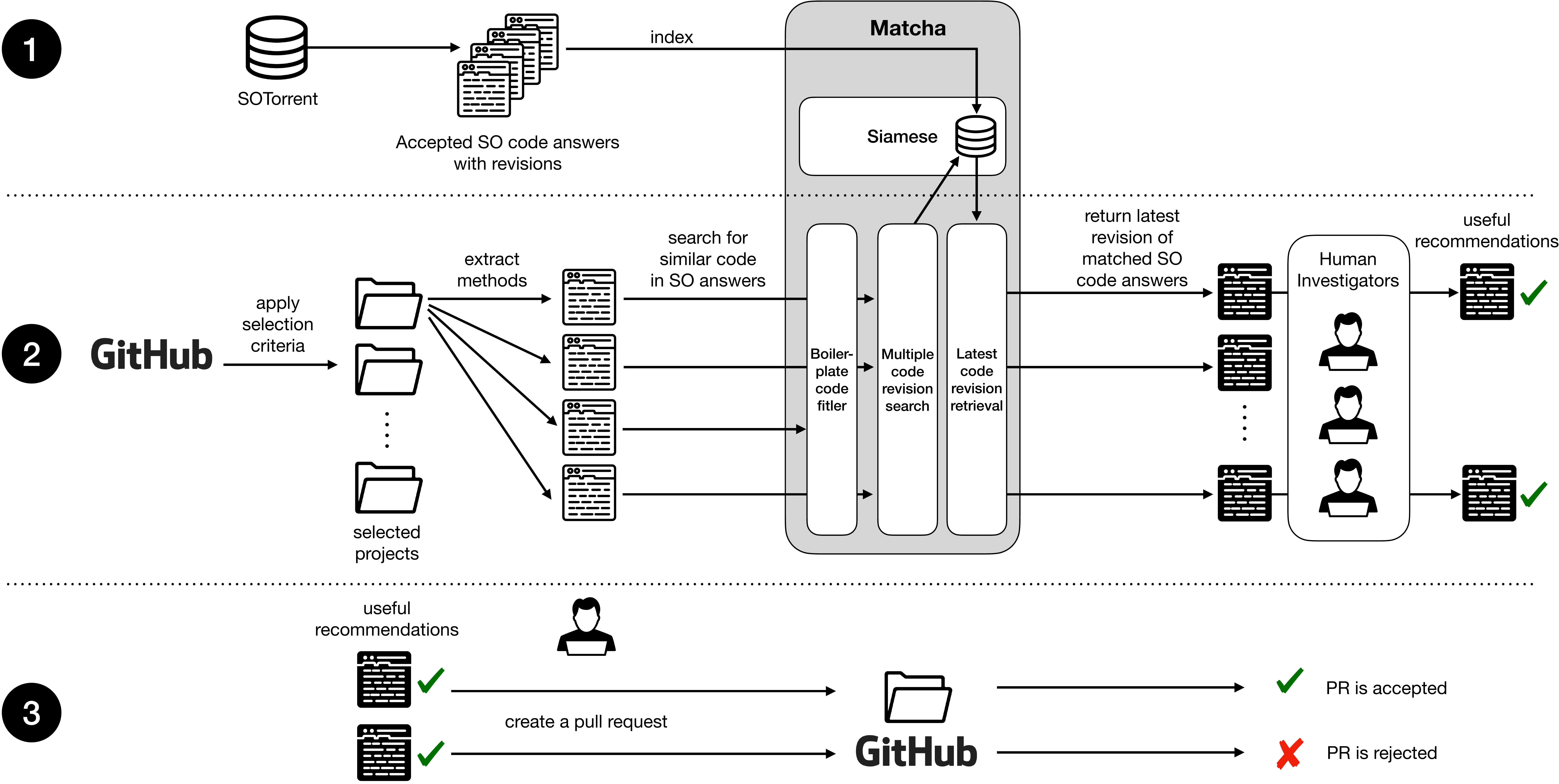}
  \caption{Overview of our exploratory study.}
  \label{fig:overview}
\end{figure*}

Figure~\ref{fig:overview} displays the overview for our exploratory study.
The study will be separated in four sequential phases (Phase 0 is not displayed in the diagram).
This section describes each phase in detail.

\color{black}
\subsection{Phase 0: \textsc{Matcha}'s Sanity Check}
\label{subsec:phase0} 
\color{black}

\color{black}
As depicted in Figure~\ref{fig:overview}, and fully explained in Section~\ref{subsec:phase2}, Siamese and the \textit{boiler plate code filter} compose the core technology underlying \textsc{Matcha}'s ability to search for similar code between a software project’s code and snippets in Stack Overflow answers.
Hence, the first step of our study must be an assessment of how well \textsc{Matcha} performs in this task.
\color{black}

\color{black}
In its original study~\cite{Ragkhitwetsagul2019}, Siamese's clone search accuracy has been evaluated using several error measures including mean average precision (MAP), mean reciprocal rank (MRR), precision-at-10 and recall on the OCD, SOCO, and BigCloneBench datasets~\cite{Ragkhitwetsagul2018,Svajlenko2014}. The evaluation results show that Siamese can search for clones with high accuracy. 
In Siamese's study, among other evaluations, the approach was assessed with two experiments that consider Stack Overflow and GitHub data. 
First, a replication study of FaCoY~\cite{Kim2018}, a code-to-code search tool, is performed.
However, the replication study only included 10 Stack Overflow code snippets as code search queries. 
Second, Siamese is used to analyse clones between the 72,365 Java code snippets in accepted answers on Stack Overflow and 16,738 GitHub Java projects for potential software license violations. 
As one can see, the Siamese clone search has not yet being directly evaluated regarding its accuracy in matching code from software projects and Stack Overflow snippets.
Moreover, in a different work~\cite{Ragkhitwetsagul2019a}, we found that the clones between Stack Overflow and software projects may be different than normal clones, such as having incomplete code snippets and a greater number of boiler plate code. 
Hence, the additional \textit{boiler plate code filter} is necessary for \textsc{Matcha}'s design, as discussed in Section~\ref{sec:design}.
Thus, it is important to evaluate Siamese in combination with the \textit{boiler plate code filter} in a comprehensive dataset as a sanity check of \textsc{Matcha}'s core underlying task.
\color{black}


For this sanity check, we will employ the established code clones ground truth between GitHub projects and Stack Overflow posts provided by the SOTorrent dataset~\cite{Baltes2018a}.
The clone pairs are established by locating source code files in a GitHub project with a URL to a Stack Overflow post in a code comment. 
The \texttt{PostReferenceGH} table contains both the URL to the Stack Overflow post and the URL to the GitHub project's file. With the latest version of SOTorrent, there is a total of 6,683,852 clone pairs recorded in the table. By filtering only Java language, there is a total of 69,885 clone pairs, which we will use as our ground truth. 

Next, we will extract the code snippets from both Stack Overflow answers and GitHub projects that appear in the ground truth.
The code snippets in GitHub projects will be extracted at a method-level granularity, while the code snippets in Stack Overflow answers will be extracted at a file-level granularity (i.e., using the whole code snippet) because some code snippets in Stack Overflow answers are not complete methods. 
We will index all the Java answers in SOTorrent, not only the answers in the ground truth,  in Siamese. This is to avoid the bias of searching only the true positives, which may affect the precision and recall of the results. 

Moreover, since existing work~\cite{Ragkhitwetsagul2018} shows that configurations can strongly affect the performance of code clone detectors and code similarity tools, and the original configurations are not always the best one, we will also perform tuning of Siamese's configurations as follows.
\color{black}
We will divide the clone pairs in the ground truth into two sets: the tuning set and the testing set, with the ratio of 20\% and 80\% of all the clone pairs accordingly. 
Next, we will run Siamese on the tuning set starting with the default configuration values of the following parameters: (1) code normalization, (2) n-gram size, (3) query reduction threshold. 
We will then perform an exhaustive search of the three parameter values to achieve the best F1 score following the method performed in our previous work~\cite{Ragkhitwetsagul2018}. 
We will call the configurations with the highest F1 score as the \textit{optimised configurations}.
Lastly, we will apply Siamese with the optimised configurations to the testing set and compare the results with the ground truth, create a confusion matrix, and measure the accuracy of the approach using measures such as precision, recall, and F1-score. 
The results will be reported in running Siamese with both original and optimised configurations.

In addition, to evaluate the effectiveness of the additional \textit{boiler-plate code filter} module on improving the accuracy of clone search, we will also evaluate \textsc{Matcha} using Siamese with the optimised configurations on the testing set before and after integrating the \textit{boiler-plate code filter} and report the comparison of the clone search accuracy (precision, recall, and F1-score) in both scenarios. Lastly, we will manually validate a sample of the clone pairs that are removed from the search results after integrating the \textit{boiler-plate code filter} to ensure the effectiveness of the filter on removing actual boiler-plate code. Interesting findings will be reported and discussed.
The evaluation performed in Phase 0 will serve as an answer to the study's RQ1: \textit{How accurate is \textsc{Matcha}’s detection of similar code snippets between software projects and Stack Overflow answers?}.
\color{black}

\subsection{Phase 1: Indexing Code Snippets from Stack Overflow's Answers on Siamese}
\label{subsec:phase1}

Upon assessing \textsc{Matcha}'s performance in matching code from software projects to snippets in Stack Overflow (Phase 0), we will move on to Phases 1-3 of our exploratory study.
In Phase 1, we will prepare \textsc{Matcha} for the later phases.
For this, we will consider the latest version of the SOTorrent dataset, as described in Section~\ref{subsec:sotorrent}.

First, we will extract the code snippets that have at least one revision from all Java accepted answers. 
Next, we import the extracted snippets into the Siamese clone search index. 
Depending on the number of code snippets to be indexed, this step may take a while.
In a previous work, the indexing of 4.8M snippets (365MLOC) in Siamese took less than a day (18 hours 13 minutes).
Our initial analysis of Java posts on SOTorrent shows that there are 3,906,637 Java posts on Stack Overflow. 
Thus, we expect the indexing time of all the Java accepted answers to take approximately one to two days.
This indexing phase occurs only once in the study and the later use of Siamese will be done by querying similar code snippets which takes a much faster time (e.g., seconds).
Furthermore, by using the incremental indexing of Siamese, the expansion of the study (e.g., add more languages) will be faster compared to the initial indexing.

\subsection{Phase 2: Recommending Code Updates from Stack Overflow Answer Edits}
\label{subsec:phase2}

In Phase 2 of our exploratory study, we will: 1) select a set of GitHub projects as an evaluation dataset, and 2) run \textsc{Matcha} for each project and assess which recommendations are useful for developers.
We detail each step as follows.


As previously mentioned, one of our assumptions for this study is that Stack Overflow answer edits may serve as a source for optimised code snippets with the potential to improve the codebase of software projects.
In this context, it is expected that projects with different levels of code quality may be benefited differently.
For instance, a project with a high-quality codebase may not have as many sub-optimal snippets, where \textsc{Matcha} would not find as many useful recommendations.
Differently, a project with a codebase containing more sub-optimal snippets might be greatly impacted by \textsc{Matcha}'s recommendations.
To evaluate this assumption, we will need a representative sample of software projects with different levels of code quality.

To search for GitHub projects, we will employ GHS (GitHub Search)~\cite{Dabic:msr2021data}, a recently published tool and dataset for searching GitHub repositories that is tailored for Mining Software Repositories studies.
The GHS dataset is composed of 735,669 repositories written in 10 programming languages.
For each project, GHS provides data regarding 25 characteristics, which can be used as filters in the search.
For this exploratory study, we will search GHS with the following filters: \texttt{Language: Java; Exclude Forks; Has Open Issues; Has Open Pull Requests.} 
We will only select projects written in \texttt{Java} due to Siamese's limitations (see Section~\ref{subsec:siamese}).
We will exclude all forks to make sure there will be no redundant projects in our dataset.
We will select projects with both open issues and open pull-requests to guarantee the projects make use of GitHub's social features to maximise the chances to have Phase~3's pull-requests properly evaluated.

For each project, GHS provides 3 popularity metrics: \texttt{Number of Stars}, \texttt{Number of Watchers} and \texttt{Number of Forks}.
GitHub popularity metrics, such as the ones provided by GHS, have been used countless times as quality proxies for GitHub projects~\cite{Zampetti2019,Gonzalez2020,Sheoran2014}.
Although isolated popularity metrics have been shown not to be the most effective method to assess a project's code quality~\cite{Munaiah2017}, we believe a combination of these metrics has the potential to yield a more trustworthy proxy.
Hence, for each quality metric provided by GHS, we compute the distribution and divide it into quartiles. 
The projects that appear above the third quartile for all metrics will be considered as having a higher quality codebase.
Similarly, the projects appearing below the first quartile for all metrics will be considered as having a lesser quality codebase.
Finally, a project that appears in between the first and third quartiles for all metrics will be considered as having a medium quality codebase.
Projects that appear in the intersection of quartiles between metrics, e.g., above the third quartile for a metric and below the first quartile for another metric, will be excluded from this study.
These criteria will provide a clear separation of projects with different levels of code quality based on the popularity metrics proxy we will employ.

\begin{figure}
  \includegraphics[width=0.9\columnwidth]{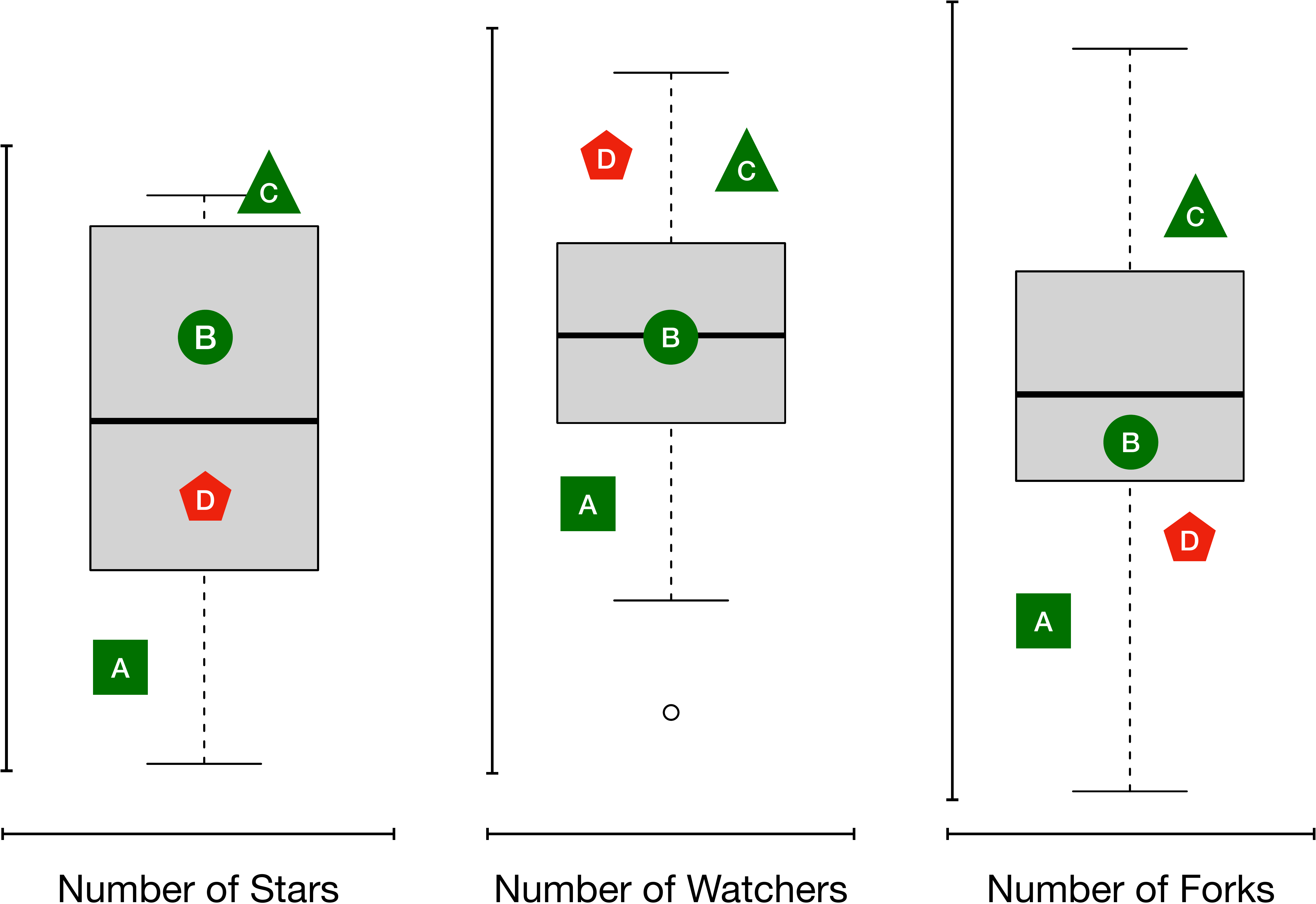}
  \caption{GitHub project selection criteria based on the distributions of number of stars, watchers, and forks.}
  \label{fig:boxplots}
\end{figure}

Figure~\ref{fig:boxplots} depicts our project selection criteria. 
Suppose the boxplots represent the distribution of the number of stars, number of watchers, and number of forks for all the projects that pass our GHS filter. There are four projects represented by four symbols:~Project A (square), Project B (circle), Project C (triangle), and Project D (pentagon). According to our project selection criteria, we will include Project A since it appears below the first quartile for all the metrics. Similarly, we will include Projects B and C since they appear between the first and third quartiles and above the third quartile for all the metrics accordingly. 
We will not include Project D because it appears in a different quartile for at least two metrics.

After applying the selection criteria discussed above, we still expect to have a population of thousands of projects, which would be infeasible to consider in our study due to the manual and qualitative methods in our evaluation.
Hence, we will compute the distribution of software projects into the three code quality tiers and take a random stratified representative sample in the 95\% confidence level.
Due to projects' constant evolution within GitHub, we will select the projects at the time of executing this exploratory study.
This way, we will have the most up to date selection of projects according to our predefined methodology.

After selecting the projects, we will run \textsc{Matcha} for each project.
\textsc{Matcha}'s execution works as follows.
First, for a given software project,~\textsc{Matcha} will extract the code for all the projects' methods and apply the \textit{boiler-plate code filter}. 
Next, for each remaining method, \textsc{Matcha} will search for similar code snippets on Siamese's index, which contains the code for Stack Overflow's accepted answers with all the versions, as shown in Phase 1.
If Siamese returns a snippet corresponding to an older version of a Stack Overflow answer, i.e., not the latest version, \textsc{Matcha} will mark this method for recommendation.
Finally, for each marked method, \textsc{Matcha} will return the latest version of the answer as a recommendation.

After running \textsc{Matcha} for each selected project and collecting all recommendations, we will enter the final step of Phase 2.
For each recommendation, we will look at the original Stack Overflow answer and perform a manual classification according to the answer edit categorisation proposed in a related paper~\cite{Baltes2020}.
Each category describes the type of edit being made in the Stack Overflow answer, such as \texttt{Optimising} and \texttt{Refactoring}.
Since \textsc{Matcha}'s recommendation is based on the latest answer edit on Stack Overflow, the category of the answer edit will also be considered the category of the recommendation provided by \textsc{Matcha}.
Consider the example provided in Section~\ref{subsec:motivating_example}.
By looking at the type of code changes performed between Listings~\ref{list:latest_code} and \ref{list:original_code}, one would categorise the edit as \texttt{Optimising}, which would also be the category of the recommendation proposed by \textsc{Matcha}.
For a full reference of the categorisation, we refer to the work by Baltes et al.~\cite{Baltes2020}.

The manual classification will be performed independently by two researchers, where they will label each recommendation into one or more categories. 
If the recommendation does not fit into any of the categories previously defined, the researcher may propose a new category.
After finishing their independent classification, the researchers will compare their labelling, and the inter-labeller reliability will be measured with Cohen's kappa coefficient.
For the recommendations with a disagreement in classification, a third researcher will intervene to provide a final classification.
After the manual classification, there will be a list of categories into which all recommendations will have been categorised.
Finally, all recommendations that perform a relevant code update, such as \texttt{Optimising} and \texttt{Refactoring}, for example, will be considered useful for developers.
The list of all categories considered useful for developers can only be known after the complete manual analysis is finished.
Hence, it is not possible, at the time of writing this paper, to present the complete list of useful categories for recommendation.

The results obtained in Phase 2 of our exploratory study will answer RQ2: \textit{To what extent are the recommendations given by \textsc{Matcha} useful for developers?}.

\subsection{Phase 3: Submitting \textsc{Matcha}'s Recommendations as Pull-Requests}
\label{subsec:phase3}

To assess \textsc{Matcha}'s potential to be adopted by software developers in their projects, we need to evaluate how the recommendations are received by developers.
For this, we will consider the results of RQ2 and collect the recommendations considered useful for developers.
Next, we will need to select a subset of recommendations to send the pull-requests.
Considering the population of all useful recommendations stratified by the projects' level of code quality and recommendations' categories (see Section~\ref{subsec:phase2}), we will extract a random stratified relevant sample at the 95\% confidence level.

\color{black}
For each recommendation in the sample, we will execute the following safeguard procedure to submit a pull-request that minimises potential issues that may be caused by our lack of knowledge regarding the project.
First, we will check whether the project provides an automated test suite.
If a test suite is not provided, the recommendation will be discarded from this evaluation phase.
Next, after incorporating the changes according to the recommended snippet, we will run the project's entire test suite.
If any test fails, the recommendation will be discarded from this evaluation phase.
For the changes that pass all tests, we will submit a pull-request that incorporates the recommended changes. 
\color{black}
The description for the pull-request will be either adapted from the Stack Overflow edit comment or created by ourselves if no edit comment was provided.

Next, we will track the status of each pull-request and avoid interfering with the review process as much as possible.
This is to mitigate any bias that may be included in the reviewing process in case we mention the exploratory study, the scientific methodology behind the pull-request or any other details outside the code change itself. 
The outcome of the pull-requests alongside the discussions between developers during review will be used to answer RQ3: \textit{To what extent are \textsc{Matcha}’s recommendations accepted in practice?}.

\section{Threats to the Validity}

\textbf{Internal validity:}
A potential threat to internal validity stems from the adoption of Siamese to search for similar code snippets in Stack Overflow accepted answers, which may contain false positives and false negatives. 
We will mitigate this threat in Phase 0 of our study by performing a sanity check of the code clone search accuracy of Siamese to evaluate if it is suitable for our approach. 
In addition, GitHub's popularity metrics (i.e. \texttt{Number of Stars}, \texttt{Number of Watchers} and \texttt{Number of Forks}) may not accurately represent a project's code quality. 
Nevertheless, we argue that a categorisation criteria that combines all three metrics and excludes projects that intersect metrics can yield a more trustworthy proxy.
Lastly, the acceptance or rejection of pull-requests may not fully represent  \textsc{Matcha}'s potential of acceptance by developers. 
We will mitigate this threat by performing a qualitative analysis of the pull-requests and the discussions between developers during review to understand the actual reason for accepting or rejecting such pull-requests.

\textbf{External validity:}
\color{black}
The ground truth data in Phase 0 is based on the code snippets in GitHub projects that have a comment pointing to the original Stack Overflow post where the code is copied from. 
The clones can potentially be biased to only clones that are copied from Stack Overflow with attributions and may not be generalised to all the clones between Stack Overflow and GitHub.
However, to the best of our knowledge, it is the only ground truth that contains a large amount of clone pairs (6,9885) with real evidence that the code has been reused from Stack Overflow to GitHub. Moreover, by relying on the code comments created by the authors of the code, we avoid the threat of manually labelling the clones by ourselves. 
\color{black}
The conclusions drawn from this study will be based on the selection of GitHub projects according to our predefined methodology detailed in Phase 2. 
It may not generalise to all GitHub projects. 
Moreover, we will only analyse code written in Java in both GitHub and Stack Overflow answers. 
Hence, the findings may not be applied to other programming languages. 
Finally, in this study, we will focus on accepted answers, where the findings may not be applicable to other types of Stack Overflow answers (e.g., newest answers, highest-voted answers).

\section{Conclusion}
This paper presents an exploratory study of recommending code improvements for sub-optimal code snippets based on the latest edit of Stack Overflow accepted answers. 
We propose an approach called \textsc{Matcha} that can search for similar code snippets in several revisions of Stack Overflow accepted answers and recommends code changes to improve the quality of a software project's code snippets. 
We plan to evaluate our approach by performing manual validation on the usefulness of the code recommendations provided by \textsc{Matcha} and by submitting pull-requests containing the useful recommendations to the GitHub projects. 
This paper also presents the methodology for selecting GitHub projects that we intend to use in our planned study.
We expect the results from the exploratory study will shed light on the potential of leveraging Stack Overflow answers for code recommendation.
\bibliographystyle{ACM-Reference-Format}
\bibliography{references}

\end{document}